\documentclass[aps,prb,twocolumn,superscriptaddress,showpacs]{revtex4-1}
\usepackage{graphicx}
\usepackage{color}
\usepackage{bm}

\begin{document}

\title{Machine learning with systematic density-functional theory calculations: Application to melting temperatures of single and binary component solids}
\author{Atsuto \surname{Seko}}
\email{seko@cms.mtl.kyoto-u.ac.jp}
\affiliation{Department of Materials Science and Engineering, Kyoto University, Kyoto 606-8501, Japan}
\author{Tomoya \surname{Maekawa}}
\affiliation{Department of Materials Science and Engineering, Kyoto University, Kyoto 606-8501, Japan}
\author{Koji \surname{Tsuda}}
\affiliation{Minato Discrete Structure Manipulation System Project, ERATO, Japan Science and Technology Agency, Sapporo 060-0814, Japan}
\affiliation{Computational Biology Research Center, National Institute of Advanced Industrial Science and Technology (AIST), Tokyo 135-0064, Japan}
\author{Isao \surname{Tanaka}}
\affiliation{Department of Materials Science and Engineering, Kyoto University, Kyoto 606-8501, Japan}
\affiliation{Center for Elements Strategy Initiative for Structure Materials (ESISM), Kyoto University, Kyoto 606-8501, Japan}
\affiliation{Nanostructures Research Laboratory, Japan Fine Ceramics Center, Nagoya 456-8587, Japan}

\date{\today}


\begin{abstract}
A combination of systematic density functional theory (DFT) calculations and machine learning techniques has a wide range of potential applications.
This study presents an application of the combination of systematic DFT calculations and regression techniques to the prediction of the melting temperature for single and binary compounds.
Here we adopt the ordinary least-squares regression (OLSR), partial least-squares regression (PLSR), support vector regression (SVR) and Gaussian process regression (GPR).
Among the four kinds of regression techniques, the SVR provides the best prediction.
The inclusion of physical properties computed by the DFT calculation to a set of predictor variables makes the prediction better.
In addition, limitation of the predictive power is shown when extrapolation from training dataset is required.
Finally, a simulation to find the highest melting temperature toward the
efficient materials design using kriging is demonstrated.
The kriging design finds the compound with the highest melting temperature much faster than random designs.
This result may stimulate the application of kriging to efficient materials design for a broad range of applications.
\end{abstract}

\maketitle

\section{Introduction}
\label{melting:introduction}
Computational material design based on data mining technique and high-throughput screening is a rapidly growing area in materials science.\cite{curtarolothe2013,MRS:Ceder_7945213,Curtarolo_PRL_2003,Fischer_NM_2006,Hautier_2012:10670628,doi:10.1146/annurev.matsci.38.060407.130217,PhysRevLett.108.068701,Greeley_2006,Castelli_C1EE02717D,Setyawan_2011,Fujimura_2013_AENM:AENM201300060}
Recent advances of computational power and techniques enable us to carry out density functional theory (DFT) calculations for a large number of compounds and crystal structures systematically. 
When the large set of DFT calculation is combined with machine learning techniques, the exploration of materials can be greatly enhanced.
Using the combination, meaningful information and pattern can be extracted from existing data to make a prediction model of a target physical property. 

In this paper, we apply the combination of systematic DFT calculations and several regression techniques to the estimation of an approximated function describing experimental melting temperatures for single and binary component solids.
So far, several theories and formulations applicable to the prediction of the melting temperature were proposed on the basis of physical considerations.
About a hundred years ago, Lindemann provided a well-known model which explains the melting temperature for single-component and simple ionic binary-component solids.\cite{Lindemann_1910}
Lindemann assumed that the critical value of the mean amplitude capable of keeping the atomic orderings in a crystal is proportional to the bond distance between atoms or ions.
Based upon the assumption and the harmonic theory, a relationship for the melting temperature $T_{\rm m}$ was derived as
\begin{equation}
T_{\rm m} = c M \Theta_{\rm D}^2 V^{2/3},
\end{equation}
where $c$, $M$, $\Theta_{\rm D}$ and $V$ denote the proportionality constant, molecular mass, Debye temperature and molar volume, respectively.
Guinea $et$ $al.$ proposed a linear relationship between the melting temperature and cohesive energy for elemental metals based on the Debye model and a binding theory of solid that they proposed.\cite{Guinea_1984}
Also for covalent crystals, a scaling theory was applied to predict their melting temperatures.\cite{PhysRevLett.29.769}
Since the theory is made for covalent crystals, it is not directly applicable to compounds with other types of chemical bondings.
Chelikowsky and Anderson demonstrated general trends of melting temperatures in some 500 AB intermetallic compounds.\cite{Chelikowsky_1987}
They found a correlation between the melting temperatures of the intermetallic compounds and those of the elemental metals A and B.

Meanwhile, a machine learning technique was applied to the prediction of the melting temperature for AB suboctet compounds recently.\cite{PhysRevB.85.104104}
They built a prediction model of the melting temperature using experimental melting temperatures of 44 suboctet AB compounds and the regularized linear regression.
They adopted only quantities of each constituent atom as predictor variables, such as atomic number, the pseudopotential radii for $s$ and $p$ orbitals and the heat of vaporization.
However, more accurate prediction models may be constructed by feeding systematic DFT results for predictors.
In addition, the use of more advanced regression technique than the linear regression used in Ref. \onlinecite{PhysRevB.85.104104} may improve the prediction.

In this study, we estimate prediction models applicable to a wide range of single and binary compounds using systematic DFT calculations and advanced regression techniques.
The set of the compounds contains a wider range of compounds than that used in the work of Ref. \onlinecite{PhysRevB.85.104104}.
We adopt four kinds of regression techniques, i.e. ordinary least-squares regression (OLSR), partial least-squares regression (PLSR),\cite{PLS_Wold_2001,PLS_VinZi_2013,PLS_Mevik_2007} support vector regression (SVR)\cite{SVR_Vapnik_1995,SVR_Vapnik_1998,Muller_SVM_914517,PRML_Bishop_2006,LIBSVM_2011} and Gaussian process regression (GPR).\cite{Rasmussen_2006}
Results by the four regression methods are compared.

Furthermore, one of the ultimate goals to use machine learning techniques is to design materials automatically.
Material design can be formulated as a complicate process to optimize target physical properties. 
Typically, the objective functions of the target physical properties cannot be defined analytically from physical laws, hence it is regarded as a ``black-box''.
Since the black-box functions can usually be supposed to be smooth, a regression function from a limited number of samples can be used as a surrogate.\cite{jones01}
In a black-box optimization technique called kriging, the measurements are designed to maximize the chance of discovering the optimal compounds.
As a case study, a simulation based on kriging for finding the compound with the highest melting temperature based on kriging is here demonstrated.

\section{Methodology}
\label{melting:method}
\subsection{Multiple linear regressions}
The common goal of regressions is to construct a model that predicts a response variable from a set of predictor variables.
The use of the multiple linear regressions allows us to attempt this goal.
In the multiple linear regressions, a linear model describes the linear relationship between a response variable and a set of predictor variables.
In the OLSR, the regression coefficients are determined by minimizing the mean-squared error for observed data.
However, when the number of predictor variables is larger than the number of observations, the OLSR cannot be applied owing to the multicollinearity.

Some approaches for avoiding the multicollinearity exist.
One is to eliminate some predictor variables.
Another is to perform the principal component analysis (PCA) of the predictor matrix and then use the principal components for the regression on the response variable.
However, it is not guaranteed that the principal components are relevant for the response variable.
The PLSR\cite{PLS_Mevik_2007,PLS_Wold_2001,PLS_VinZi_2013} is an extension of the OLSR and combines features of OLSR and PCA.
By contrast to the PCA, the PLSR extracts underlying factors from predictor variables that are relevant for the response variable.
They are called latent variables and described by linear combinations of the predictor variables.
By using a small number of latent variables, it is possible to build a linear prediction model from a large number of the predictor variables with avoiding the multicollinearity.

The PLSR with a single response variable finds a set of the latent variables that performs a simultaneous decomposition of $\bm{X}$ and $\bm{y}$, where $\bm{X}$ and $\bm{y}$ are $(N \times m)$ predictor matrix containing $m$ predictor variables for $N$ observed data and $N$-vector of the response variable, respectively.
In a PLSR model with $H$ latent variables, the predictor matrix is decomposed as
\begin{equation}
\bm{X} = \bm{T} \bm{P}^\top
\end{equation}
in the same fashion as the PCA, where $\bm{T}$ and $\bm{P}$ denote $(N \times H)$ score matrix and $(m \times H)$ loading matrix for $\bm{X}$, respectively.
The score matrix is a collection of the latent vectors and expressed as $\bm{T} = [\bm{t}_1, \cdots, \bm{t}_H]$, where $\bm{t}_h$ is $h$-th latent vector.
The loading matrix is not orthogonal in the PLSR opposite to the PCA.
Similar to the predictor matrix, the response variable is also decomposed as
\begin{equation}
\bm{y} = \bm{T} \bm{q},
\end{equation}
where $\bm{q}$ is a vector with $H$ components equivalent to the product of a diagonal matrix and loadings for $\bm{y}$.

Scores and loadings are obtained by an iterative procedure.
The detailed procedure is shown in Table \ref{melting:pls_algorithm}.
Firstly, a pair of $\bm{t}_1$ and weight vector $\bm{w}_1$ with the relationship of $\bm{t}_1 = \bm{X} \bm{w}_1$ are determined with the constraint that $\bm{t}_1^\top \bm{y}$ is maximized.
Once the first latent vector and loadings are found, it is subtracted from $\bm{X}$ and $\bm{y}$.
This is repeated until $H$-th latent vector, weight vector and loadings are found.
Using the weight matrix $\bm{W} = [\bm{w}_1, \cdots, \bm{w}_H]$ obtained by the iterative procedure, the regression model is written as
\begin{equation}
\bm{y} = \bm{X} \bm{W}^{*} \bm{q} =  \bm{X} \bm{b} ,
\end{equation}
where $\bm{W}^*$ has the equality of $\bm{W}^{*} = \bm{W}\left( \bm{P}^\top \bm{W} \right)^{-1}$. 
Consequently, the linear regression coefficient vector $\bm{b}$ corresponds to $\bm{W}^{*} \bm{q}$.

\begin{table}[tbp]
\caption{
Algorithm for building PLSR model with a single response variable.
$\bm{p}_h$, $q_h$ are $h$-th components of $\bm{P}$ and $\bm{q}$, respectively.
$\bm{E}_h$ and $\bm{f}_h$ are the residual for the predictor matrix and that for the response variable, respectively.
}
\label{melting:pls_algorithm}
\begin{ruledtabular}
\begin{tabular}{ll}
Input: $\bm{E}_0 = \bm{X}$, $\bm{f}_0 = \bm{y}$ \\
Output: $\bm{W}$, $\bm{q}$, $\bm{T}$, $\bm{P}$ \\
\hspace{3mm} for all $h = 1, \ldots, H$ do  \\
\hspace{6mm} Step 1: $\bm{w}_h = \bm{E}^{\rm{T}}_{h-1} \bm{f}_{h-1} / || \bm{E}^{\rm{T}}_{h-1} \bm{f}_{h-1}|| $ \\
\hspace{6mm} Step 2: $\bm{t}_h = \bm{E}_{h-1} \bm{w}_{h} / (\bm{w}^{\rm{T}}_{h} \bm{w}_{h}) $ \\
\hspace{6mm} Step 3: $q_h = \bm{f}^{\rm T}_{h-1} \bm{t}_{h} / (\bm{t}^{\rm{T}}_{h} \bm{t}_{h}) $ \\
\hspace{6mm} Step 4: $\bm{p}_h = \bm{E}^{\rm T}_{h-1} \bm{t}_{h} / (\bm{t}^{\rm{T}}_{h} \bm{t}_{h}) $ \\
\hspace{6mm} Step 5: $\bm{E}_h = \bm{E}_{h-1} - \bm{t}_h \bm{p}^{\rm T}_h$ \\
\hspace{6mm} Step 6: $\bm{f}_h = \bm{f}_{h-1} - q_h \bm{t}_h$ \\
\hspace{3mm} end for 
\end{tabular}
\end{ruledtabular}
\end{table}

\subsection{Nonlinear regressions}
\subsubsection{Support vector regression (SVR)}
To approximate complex response functions, many frameworks 
beyond the linear regression have been proposed.
SVR is a regression version of support vector machines 
that constructs a nonlinear regression function based on a kernel function. 

Consider a set of $N$ training data $\left\{ (\bm{x}_1, y_1), \cdots, (\bm{x}_N, y_N) \right\}$, where $\bm{x}_i$ and $y_i$ denote a vector of predictor variables and the response variable.
Let $\bm{w}$ and $b$ denote the weight vector 
and the bias parameter, respectively.
In $\epsilon$-SVR, the response function $f(\bm{x})$ is modeled as
\begin{equation}
f(\bm{x}) = \bm{w}^\top \phi(\bm{x}) + b,
\end{equation}
where $\phi(\bm{x})$ maps $\bm{x}$ into a higher-dimensional space.
We define $\phi(\bm{x})$ in an implicit form using a kernel function 
$k(\bm{x}, \bm{x}^\prime)$ as
\[
k(\bm{x},\bm{x}^\prime) = \phi(\bm{x})^\top \phi(\bm{x}^\prime).
\]
It has been proven that a mapping $\phi$ exists if and only if the
kernel function is positive semidefinite.
A popular choice of $k$ includes the Gaussian kernel and the polynomial kernels.

Introducing non-negative slack variables $\xi_i$ and $\xi_i^*$ to allow for some errors, the optimization problem is stated as
\begin{eqnarray}
\min_{\bm{w},b,\bm{\xi},\bm{\xi}^*} \frac{1}{2} \bm{w}^\top \bm{w} + C \sum_{i=1}^{N} \left( \xi_i + \xi_i^* \right) \nonumber
\end{eqnarray}
\begin{equation}
\mbox{\rm subject to  }
\left\{
\begin{array}{l}
\bm{w}^\top \phi(\bm{x_i}) + b - y_i \leq \epsilon + \xi_i,\\
y_i - \bm{w}^\top \phi(\bm{x_i}) - b \leq \epsilon + \xi_i^* \\
\xi_i, \xi_i^* \geq 0, i = 1, \cdots, N
\end{array}
\right.
\end{equation}
where $C$ denotes a positive regularization parameter.
This corresponds to dealing with a so-called $\epsilon$-insensitive loss function $|\xi|_\epsilon$ expressed by
\begin{equation}
|\xi|_\epsilon = 
\left\{
\begin{array}{ll}
0 & \mbox{if } |\xi| \leq \epsilon \\
|\xi| - \epsilon &\mbox{otherwise.}
\end{array}
\right.
\end{equation}
The $\epsilon$-insensitive loss function ignores errors less than $\epsilon$.

The optimization problem can be solved more easily in its dual formulation.
In general, a standard dualization method based on 
Lagrange multipliers is applied to the optimization.
The dual problem is stated as
\begin{eqnarray}
\min_{\bm{\alpha},\bm{\alpha}^*} & & \frac{1}{2} \left(\bm{\alpha}-\bm{\alpha}^* \right)^\top \bm{K} (\bm{\alpha} - \bm{\alpha}^*) \nonumber \\
& + &\epsilon \sum_{i=1}^{N} \left( \alpha_i + \alpha_i^* \right) \nonumber + \sum_{i=1}^{N} y_i \left( \alpha_i - \alpha_i^* \right) \nonumber
\end{eqnarray}
\begin{eqnarray}
\mbox{\rm subject to   }
\left\{
\begin{array}{l}
\bm{e}^\top \left(\bm{\alpha} - \bm{\alpha}^*\right) = 0,\\
0 \leq \alpha_i, \alpha_i^* \leq C, i = 1, \ldots, N
\end{array}
\right.
\end{eqnarray}
where $\bm{e}$ is the vector of all ones, and $\bm{\alpha}$ and $\bm{\alpha}^*$ are Lagrange multipliers.
Here, $\bm{K}$ is called a kernel matrix whose $(i,j)$ component is 
$k(\bm{x}_i,\bm{x}_j)$.
Using the obtained $\bm{\alpha}$ and $\bm{\alpha}^*$, 
the response function is written as
\begin{equation}
f(\bm{x}) = \sum_{i=1}^{N} \left(- \alpha_i + \alpha_i^* \right) k(\bm{x}_i, \bm{x}) + b.
\end{equation}

\subsubsection{Gaussian process regression (GPR)}
The GPR is one of Bayesian regression techniques and has been successfully employed to solve nonlinear estimation problems.
A Gaussian process is a generalization of the multivariate Gaussian probability distribution.
The prediction $f(\bm{x}_*)$ at a point $\bm{x}_*$ and its variance $v(f_*)$ are described by using the Gaussian kernel function as follows,
\begin{equation}
k\left(\bm{x}_i, \bm{x}_j \right) = \exp \left( - \frac{|\bm{x}_i - \bm{x}_j|^2}{2 \sigma^2} \right).
\end{equation}
When the prior distribution has a variance of $\sigma^2$, the prediction is given as
\begin{equation}
f(\bm{x}_*) = {\bm k}_*^\top  (\bm{K} + \sigma^2 \bm{I})^{-1} \bm{y},
\end{equation}
where ${\bm k}_* = \left[ k(\bm{x}_1, \bm{x}_*), \cdots, k(\bm{x}_N, \bm{x}_*) \right] ^\top$ is the vector of kernel values between $\bm{x}_*$ and the training examples, and $\bm{I}$ is the unit matrix. 
The prediction variance is described as
\begin{equation}
v(f_*) = k(\bm{x}_*,\bm{x}_*) - {\bm k}_*^\top (\bm{K} + \sigma^2 \bm{I})^{-1} {\bm k}_*.
\end{equation}

\subsection{Kriging}
\begin{figure}
\begin{tabular}{c}
\includegraphics[width=0.7\linewidth]{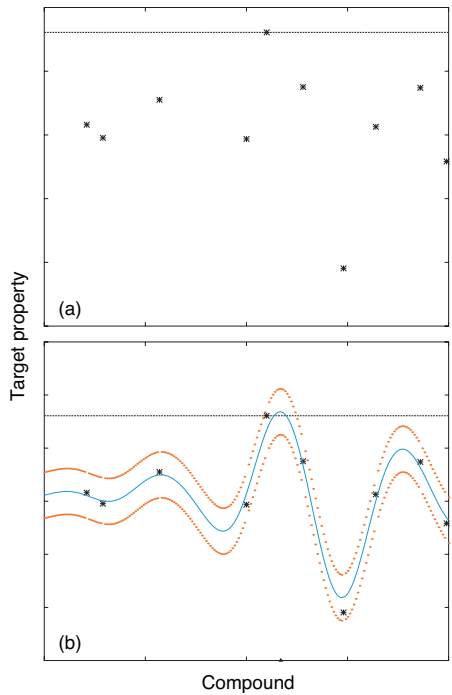}
\end{tabular}
\caption{
Illustration of kriging. 
(a) A typical situation where several samples are available.
The current best target property is shown as a horizontal line.
(b) The GPR is applied to the available samples.
The prediction of the target property by the GPR is shown by the blue line.
The probability distribution of the target property at all possible compounds is also shown by orange closed circles.
}
\label{fig:kri}
\end{figure}

Kriging is built on the Gaussian processes.
Figure \ref{fig:kri} (a) shows a typical situation where several samples are available. 
In kriging, we search for a next sampling point where the chance of getting beyond the current best target property is optimal. 
To this aim, a Bayesian regression method such as a Gaussian process is applied, and the probability distribution of the score at all possible parameter values is obtained as illustrated in Fig. \ref{fig:kri} (b).
Then, the next sampling point is determined as the one with the highest probability of improvement.

We here apply the kriging to find the compound with the highest melting temperature from a pool of compounds.
The procedure used in this study is organized as follows.
\begin{enumerate}
\item An initial training set is firstly prepared by choosing compounds randomly.
\item Then a compound is selected using based on the GPR.
The compound is chosen as the one with the largest probability of getting beyond the current best value $f_{cur}$.
Since the probability is a monotonically increasing function of the z-score,
\begin{equation}
z({\bm x}_*) = (f(\bm{x}_*) - f_{cur})/\sqrt{v(\bm{x}_*)},
\end{equation}
the compound with the highest z-score is chosen from the pool of unobserved materials.

\item The melting temperature of the selected compound is observed.
\item The selected compound is added into the training data set.
Then the simulation goes back to step (2).
\end{enumerate}

Steps (2)-(4) are repeated until all data of melting temperatures are included in the training set.

\section{Results and discussion}
\label{melting:result}
\subsection{Data set}
Prediction models are built from a data set containing experimental melting temperatures and predictors for 248 compounds. 
The melting temperatures of the 248 compounds range from room temperature to 3273 K as shown in Appendix.
The set of compounds do not contain transition metals to avoid complexity in the DFT calculations.

In order to make prediction models, the compounds are characterized by elemental information and simple physical properties of the compounds.
These features are used as predictor variables.
A key factor for constructing accurate prediction models is to supply good predictor variables.
We here adopted (1) cohesive energy, $E_{\rm coh}$, (2) bulk modulus, $B$, (3) volume, $V$, and (4) nearest-neighbor pair distance, $r_{\rm NN}$ as the physical properties of compounds as shown in Table \ref{melting:dft_predictor}.
They are systematically obtained by the DFT calculation.
Besides the physical properties computed by the DFT calculation, ten kinds of elemental information are adopted, i.e.,
(1) atomic number of elements A and B, $Z_{\rm A}$, $Z_{\rm B}$, 
(2) atomic mass of elements A and B, $m_{\rm A}$, $m_{\rm B}$, 
(3) number of valence electrons of elements A and B, $n_{\rm A}$, $n_{\rm B}$, 
(4) periods in periodic table of elements A and B, $p_{\rm A}$, $p_{\rm B}$, 
(5) groups in periodic table of elements A and B, $g_{\rm A}$, $g_{\rm B}$, 
(6) van der Waals radius of elements A and B, $r_{\rm A}^{\rm vdw}$, $r_{\rm B}^{\rm vdw}$, 
(7) covalent radius of elements A and B, $r_{\rm A}^{\rm cov}$, $r_{\rm B}^{\rm cov}$, 
(8) Pauling electronegativity of elements A and B, $\chi_{\rm A}$, $\chi_{\rm B}$, 
(9) first ionization energy of elements A and B, $I_{\rm A}$, $I_{\rm B}$, 
and (10) compositions of A$_x$B$_y$ compound for elements A and B, $c_{\rm A}= x/(x+y)$, $c_{\rm B}=y/(x+y)$.
In practice, symmetric forms of the elemental information are introduced so that predictor variables become symmetric for the exchange of atomic species in binary compounds.
The symmetric forms are shown in Table \ref{melting:symmetric_form}.
As a result, the total number of predictor variables is 23.

\begin{table}[tbp]
\caption{
Physical properties of compounds adopted as predictor variables.
They are computed by the DFT calculation.
}
\label{melting:dft_predictor}
\begin{ruledtabular}
\begin{tabular}{lc}
Physical property & \\
\hline
Volume & $V$ ($x_1$) \\
Nearest-neighbor pair distance & $r_{\rm NN}$ ($x_2$) \\
Cohesive energy & $E_{\rm coh}$ ($x_3$) \\
Bulk modulus & $B$ ($x_4$) \\
\end{tabular}
\end{ruledtabular}
\end{table}

\begingroup
\squeezetable
\begin{table}[tbp]
\caption{
Symmetric forms of predictor variables composed of elemental information.
The elemental information are taken from Ref. \onlinecite{CRC-Handbook-Chemistry-Physics}.
}
\label{melting:symmetric_form}
\begin{ruledtabular}
\begin{tabular}{lcc}
& Sum form & Product form \\
\hline
Composition, $c$ & & $c_{\rm A} c_{\rm B}$ ($x_5$) \\
Atomic number, $Z$ & $Z_{\rm A} + Z_{\rm B}$ ($x_6$) & $Z_{\rm A} Z_{\rm B}$ ($x_7$) \\
Atomic mass, $m$ & $m_{\rm A} + m_{\rm B}$ ($x_8$) & $m_{\rm A} m_{\rm B}$ ($x_9$) \\
Number of valence electrons, $n$ & $n_{\rm A} + n_{\rm B}$ ($x_{10}$) & $n_{\rm A} n_{\rm B}$ ($x_{11}$) \\
Group, $g$ & $g_{\rm A} + g_{\rm B}$ ($x_{12}$) & $g_{\rm A} g_{\rm B}$ ($x_{13}$) \\
Period, $p$ & $p_{\rm A} + p_{\rm B}$ ($x_{14}$) & $p_{\rm A} p_{\rm B}$ ($x_{15}$) \\
van der Waals radius, $r^{\rm vdw}$ & $r^{\rm vdw}_{\rm A} + r^{\rm vdw}_{\rm B}$ ($x_{16}$) & $r^{\rm vdw}_{\rm A} r^{\rm vdw}_{\rm B}$ ($x_{17}$) \\
Covalent radius, $r^{\rm cov}$ & $r^{\rm cov}_{\rm A} + r^{\rm cov}_{\rm B}$ ($x_{18}$) & $r^{\rm cov}_{\rm A} r^{\rm cov}_{\rm B}$ ($x_{19}$) \\
Electronegativity, $\chi$ & $\chi_{\rm A} + \chi_{\rm B}$ ($x_{20}$) & $\chi_{\rm A} \chi_{\rm B}$ ($x_{21}$) \\
First ionization energy, $I$ & $I_{\rm A} + I_{\rm B}$ ($x_{22}$) & $I_{\rm A} I_{\rm B}$ ($x_{23}$) 
\end{tabular}
\end{ruledtabular}
\end{table}
\endgroup

The DFT computation of physical properties requires the crystal structure for each compound.
Candidates for the crystal structure are taken from the Inorganic Crystal Structure Database (ICSD).
When ICSD database has a unique crystal structure for a compound, the DFT calculation is carried out by using the unique crystal structure.
When ICSD database contains multiple crystal structures for a compound, DFT calculations for all the crystal structures are performed.
The crystal structure with the lowest energy is then adopted for computing the physical properties.

The cohesive energy is computed by the DFT calculation using the formula normalized by the total number of atoms, expressed as
\begin{equation}
E_{\rm coh} = \frac{(N_{\rm A} E_{\rm A}^{\rm atom} + N_{\rm B} E_{\rm B}^{\rm atom}) - E^{\rm bulk}}{N_{\rm A} + N_{\rm B}}, 
\end{equation}
where $N_{\rm A}$ and $N_{\rm B}$ denote the numbers of atoms A and B included in the simulation cell, respectively.
$E^{\rm bulk}$ is the total energy of compound at the equilibrium volume.
$E_{\rm A}^{\rm atom}$ and $E_{\rm B}^{\rm atom}$ are the atomic energies of A and B, respectively.
Here, the energy of an isolated atom in a large cell ($=10$ \AA $\times 10$ \AA $\times 10$ \AA) is regarded as the atomic energy.
The bulk modulus $B$ is evaluated using the formula of 
\begin{equation}
B = - V_0 \frac{\partial P}{\partial V} = -V_0 \frac{P_1-P_0}{V_1-V_0},
\end{equation}
where $V_0$ and $V_1$ denotes the equilibrium volume and the volume that is slightly different from the equilibrium volume, respectively.
$P_0$ and $P_1$ are the pressure at volumes $V_0$ and $V_1$, respectively.

DFT calculations are performed by the projector augmented-wave (PAW) method\cite{PAW1,PAW2} within the generalized gradient approximation (GGA)\cite{GGA:PBE96} as implemented in the VASP code.\cite{VASP1,VASP2}
The total energies converge to less than 10$^{-2}$ meV.
The atomic positions and lattice constants are relaxed until the residual forces become less than $10^{-3}$ eV$/$\AA.

\subsection{Regressions}
\label{melting:regressions}
Regressions are carried out using two kinds of predictor variable set.
Predictor set (1) is composed only of symmetric predictor variables of elemental information as listed in Table \ref{melting:symmetric_form}.
Predictor set (1) contains no information obtained by the DFT calculation.
Predictor set (2) is composed of symmetric predictor variables of elemental information and physical properties of compounds computed by the DFT calculation.

In order to estimate the prediction error, we divide the data set into training and test data.
A randomly-selected quarter of the data set and the rest of the data set are regarded as the test and training data, respectively.
This is repeated thirty times and then averages of 10-fold cross validation (CV) scores and the root-mean-square (RMS) errors between predicted and experimental melting temperatures for test data are evaluated.

We first perform the OLSR for building prediction models.
Table \ref{melting:prediction_error} shows the CV scores of the OLSR models. 
When using predictor sets (1) and (2), we construct prediction models with the CV scores of 473 K and 293 K, respectively.
The prediction is improved by considering physical properties of compounds computed by the DFT calculation as predictor variables.
We then perform the PLSR using two kinds of predictor variable sets.
The PLSR is performed using {\it R package}.\cite{PLS_Mevik_2007}
The accuracy of the PLSR is mainly controlled by the number of the latent factors.
The CV scores converge at the number of latent factors of 18 and 20 using predictor set (1) and using predictor set (2), respectively.
Table \ref{melting:prediction_error} shows the CV scores of the optimized PLSR models. 
When predictor sets (1) and (2) are used, we construct prediction models with the CV scores of 476 K and 291 K, respectively.
They are almost the same as the CV scores of OLSR models because the OLSR models are made with less uncertainty.

\begin{table}[tbp]
\caption{
CV scores and RMS errors for test data in OLSR, PLSR, SVR and GPR.
}
\label{melting:prediction_error}
\begin{ruledtabular}
\begin{tabular}{ccc}
& CV score (K) & RMS error for test data (K)\\
\hline
Predictor set (1) & & \\
OLSR & 473 & 472 \\
PLSR & 476 & 476 \\
SVR & 376 & 364 \\
GPR & 492 & 481 \\
\hline
Predictor set (2) & & \\
OLSR & 293 & 306 \\
PLSR & 291 & 305 \\
SVR & 265 & 262 \\
GPR & 334 & 306 
\end{tabular}
\end{ruledtabular}
\end{table}

\begin{figure}[tbp]
\begin{center}
\includegraphics[width=\linewidth,clip]{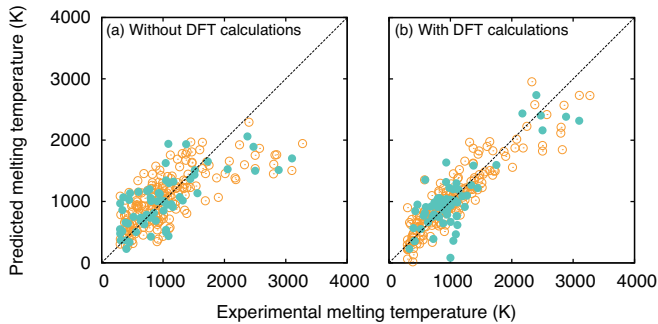} 
\caption{
Melting temperature for 248 compounds predicted by the OLSR performed by (a) predictor set (1) and (b) predictor set (2).
This is obtained from one of the thirty kinds of random divisions of the data set into training and test data.
Melting temperatures of training and test data are shown by open and closed circles, respectively.
On the broken line, experimental and predicted melting temperatures are exactly the same.
}
\label{melting:ols_prediction}
\end{center}
\end{figure}

The RMS errors for test data of the OLSR models using predictor sets (1) and (2) are 472 and 306 K, respectively, which are almost the same as the CV scores.
Also in the case of the PLSR, the prediction errors are almost the same as the CV scores.
Figure \ref{melting:ols_prediction} shows relationships of predicted and experimental melting temperatures using predictor sets (1) and (2) in the OLSR.
This is obtained from one of the thirty kinds of random divisions of the data set.
As can be seen in Fig. \ref{melting:ols_prediction}, the prediction errors for the training and test data are comparable in the OLSR models using both predictor sets (1) and (2) since the CV score and RMS error for test data are also comparable.
The deviation from the straight line, on which the experimental and predicted melting temperatures are equal, in the OLSR model using predictor set (1) is larger than that in the OLSR model using predictor set (2), corresponding to values of the CV score and the RMS error for test data.

\begin{figure}[tbp]
\begin{center}
\includegraphics[width=\linewidth,clip]{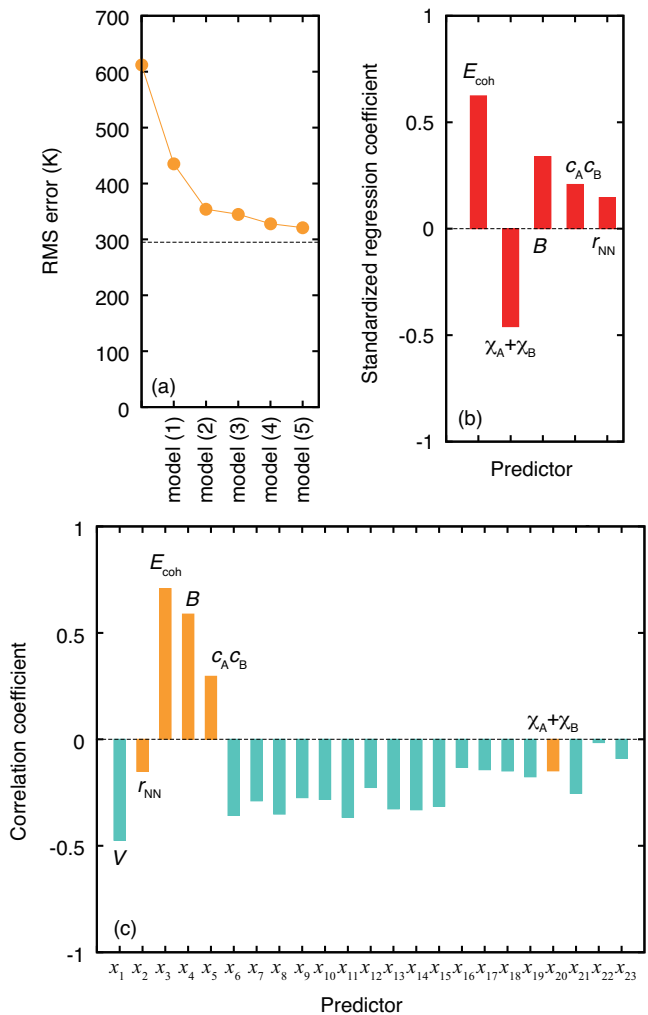} 
\caption{
(a) RMS error of five prediction models with up to five predictors selected during the stepwise regression method.
The predictor sets of models (1)-(5) are composed of $E_{\rm coh}$, $\{E_{\rm coh},\chi_{\rm A} + \chi_{\rm B}\}$, $\{E_{\rm coh},\chi_{\rm A} + \chi_{\rm B}, B\}$, $\{E_{\rm coh},\chi_{\rm A} + \chi_{\rm B}, B, c_{\rm A} c_{\rm B} \}$ and $\{E_{\rm coh},\chi_{\rm A} + \chi_{\rm B}, B, c_{\rm A} c_{\rm B},r_{\rm NN} \}$.
(b) Standardized regression coefficients of the prediction model (5) for the five predictors.
(c) Correlation coefficients between the melting temperature and predictors.
Orange solid bars show the correlation coefficients for the predictors of model (5).
}
\label{melting:ols_physics}
\end{center}
\end{figure}

To find important predictors for explaining the melting temperature, a selection of predictors within the OLSR using predictor set (2) is then carried out.
We adopt a stepwise regression method with the bidirectorial elimination\cite{Venables_MASS} based on the minimization of the Akaike's information criterion (AIC).\cite{akaike1973information}
As a result, the best prediction model with the minimum AIC is composed of ten predictors and has the RMS error of 295 K.
Figure \ref{melting:ols_physics} (a) shows the RMS errors for prediction models only with up to five predictors obtained during the stepwise regression.
The prediction model with five predictors shows the RMS error of 320 K which is close to that of the best prediction model.
The selected five predictors are $E_{\rm coh}$, $\chi_{\rm A} + \chi_{\rm B}$, $B$, $c_{\rm A} c_{\rm B}$ and $r_{\rm NN}$.
Three of the five predictors are physical properties of compounds computed by the DFT calculation.
Figure \ref{melting:ols_physics} (b) shows the standardized regression coefficients of the prediction model with the five predictors.
The earlier the predictors are selected by the stepwise regression, the larger the absolute value of the standardized regression coefficients for the predictors are.
The absolute value of the standardized regression coefficient for $E_{\rm coh}$, which is the first selected by the stepwise regression, is the largest among the coefficients for the five predictors, hence it can be considered that $E_{\rm coh}$ contributes the most to the prediction of the melting temperature.
The importance of the predictors for explaining the melting temperature can be seen in the correlations between the melting temperature and predictors.
Figure \ref{melting:ols_physics} (c) shows the correlation coefficients between the melting temperature and predictors.
The correlation coefficients of $E_{\rm coh}$ and $B$, which are selected by the stepwise regression, are positively large.
On the other hand, $V$ is not selected by the stepwise regression in spite of its negatively-large correlation coefficient.
This may be ascribed by the fact that the correlations between $V$ and the other physical properties computed by the DFT calculation are large.

Next, we perform the SVR and GPR using predictor sets (1) and (2).
The SVR and GPR are performed using {\it R package}.\cite{e1071,kernlab}
The Gaussian kernel is adopted as the kernel function in the SVR.
The SVR with the Gaussian kernel has two parameters which control the accuracy of the prediction model, i.e., the variance of the Gaussian kernel and the regularization parameter.
Therefore, the two parameters are optimized based on the minimization of the CV score.
Candidates of them are set to $10^{-3}$, $10^{-2}$, $10^{-1}$, $10^0$, $10^1$, $10^2$ and $10^3$.
By performing regressions for all combinations of the candidates, the optimal values of the two parameters are determined.

Table \ref{melting:prediction_error} shows the CV scores of the optimized SVR and GPR models. 
Using the SVR, we get prediction models with the CV scores of 376 K and 265 K using predictor sets (1) and (2), respectively.
Using the GPR, prediction models with the CV scores of 492 K and 334 K are obtained using predictor sets (1) and (2), respectively.
As is the case in the OLSR, the prediction of the melting temperature is improved by considering physical properties of compounds computed by the DFT calculation as predictors.
In addition, when using predictor set (1), the SVR model is the best among the four kinds of regression models.
On the other hand, when using predictor set (2), the use of the SVR does not improve the prediction well compared to the linear regressions.

\begin{figure}[tbp]
\begin{center}
\includegraphics[width=\linewidth,clip]{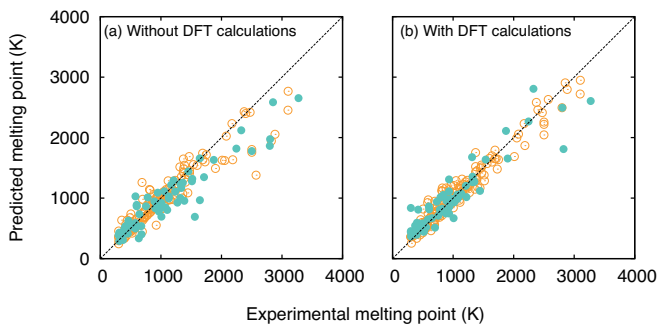} 
\caption{
The same as in Figure \ref{melting:ols_prediction} but for SVR models.
}
\label{melting:SVR_prediction}
\end{center}
\end{figure}

\begin{figure}[tbp]
\begin{center}
\includegraphics[width=\linewidth,clip]{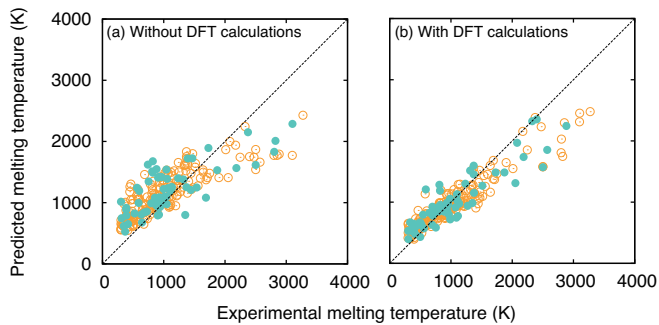} 
\caption{
The same as in Figure \ref{melting:ols_prediction} but for GPR models.
}
\label{melting:GPR_prediction}
\end{center}
\end{figure}

Figure \ref{melting:SVR_prediction} and \ref{melting:GPR_prediction} show relationships of predicted and experimental melting temperatures in the SVR and GPR, respectively.
They are obtained from one of the thirty kinds of random divisions of the data set, the same as those in the OLSR.
Then the RMS errors for test data are also estimated.
The RMS errors of SVR models using predictor sets (1) and (2) are 364 and 262 K, respectively, which are very close to the CV scores.
The RMS errors of GPR models using predictor sets (1) and (2) are 481 and 306 K, respectively, which are also close to the CV scores.
Among the four kinds of regression techniques, the SVR provides the prediction models with the best CV scores and RMS errors.
This is consistent with the fact that nonlinear regressions are widely accepted to be useful for estimating complex response functions.

\subsection{Prediction}
In this section, we examine the predictive power of the melting temperature of compounds which are missing in the dataset of Table \ref{melting:data} (hereafter called dataset I).
Some nitrides are known to decompose releasing nitrogen gas at a temperature below the melting point under the ambient pressure.
The decomposition temperature is shown instead of the melting temperature in some databases.
Elemental carbon is another example whose melting temperature under the ambient pressure is not well established by experiments.
A series of nitrides and elements of Group 14 (carbon group) are therefore selected for the targets of the prediction. 

\begin{figure}[tbp]
\begin{center}
\includegraphics[width=\linewidth,clip]{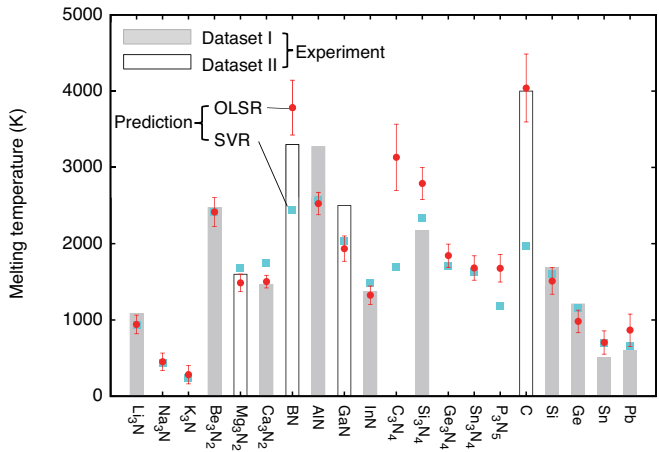} 
\caption{
Melting temperatures of nitrides and Group 14 elements predicted by the SVR (blue closed squares) and the OLSR with ten predictors optimized by the stepwise method (red closed circles) along with experimental melting temperatures (gray closed bars) in dataset I.
The error bars indicate 95\% confidence intervals in the OLSR model.
Open bars show melting temperatures of Mg$_3$N$_2$, BN, GaN and C obtained by extrapolation using the solid-liquid phase boundaries in pressure-temperature phase diagrams (dataset II).
}
\label{melting:nitride}
\end{center}
\end{figure}

Figure \ref{melting:nitride} shows melting temperatures for nitrides and Group 14 elements predicted with dataset I by the SVR model and the OLSR model, which lead to RMS errors for test data of 262 K and 295 K. 
Ten predictors are optimized by the stepwise method in the OLSR model.
The error bars shown in Fig. \ref{melting:nitride} correspond to 95\% confidence intervals in the OLSR model.
The melting temperatures predicted by SVR and OLSR models do not differ so much for most of compounds included in the dataset I.
They are also close to experimental melting temperatures.
The largest error can be found for AlN.
The reason for the poor prediction may be ascribed to an experimental error rather than problems in the prediction model, since the experimental data in literature is widely scattered.
It is 3273.15 K in dataset I, while other databases report 2473.15 K\cite{MAWE:MAWE19970280206} and 3473 K.\cite{ambacher1998growth}

Meanwhile, missing compounds in dataset I can be classified into two groups according to the width of the error bar in Fig. \ref{melting:nitride}.
For compounds with narrow error bars, the predictions by SVR and OLSR models are nearly the same, which is similar to those compounds in dataset I.
In such a compound, the melting temperature is expected to be predictable with the accuracy comparable to that for compounds in dataset I.
We collected experimental melting temperatures of compounds that are not included in dataset I and made a new dataset II.
They are estimated from an extrapolation of experimental solid-liquid phase boundary to the ambient pressure in a pressure-temperature phase diagram. 

As can be seen in Fig. \ref{melting:nitride}, the melting temperatures predicted by SVR and OLSR models using dataset I agree well with the experimental data in dataset II when the error bar of the prediction is narrow, as for Mg$_3$N$_2$. 
On the other hand, the prediction is less reliable for compounds with wide error bars such as C and BN.
In contrast to compounds with narrow error bars, the melting temperature predicted by the OLSR model differs greatly from those predicted by the SVR model.
The prediction with the wide error bars requires an extrapolation from the dataset I.
As demonstrated in Sec \ref{melting:regressions}, both of the cohesive energies and bulk moduli of the compounds are important predictors in the OLSR model. 
Since both C and BN have larger cohesive energy and bulk modulus than those of compounds in the dataset I, their melting temperatures need to be predicted by the extrapolation. 
Hence, the predictive power for these compounds becomes poor. 
Inclusion of these new data into the training dataset should decrease the uncertainty of the prediction models, thereby improving the predictive power drastically.

\subsection{Kriging}
\begin{figure}[tbp]
\begin{center}
\includegraphics[width=0.8\linewidth,clip]{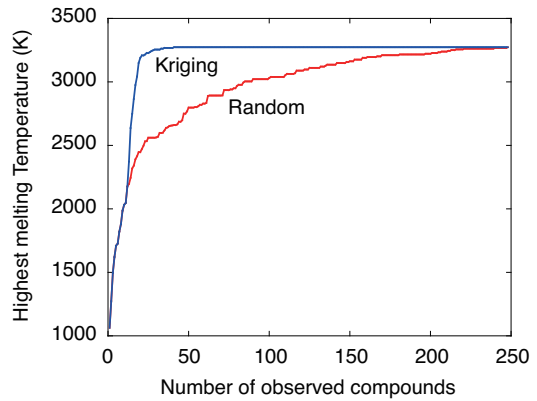} 
\caption{
Highest melting temperature among the observed compounds in simulations for finding the compound with the highest melting temperature based on the kriging and random compound selections.
}
\label{melting:kriging}
\end{center}
\end{figure}

Finally, we perform a simulation for finding the compound with the highest melting temperature using the kriging.
Here we start the kriging from a data set of 12 compounds.
For comparison, a simulation based on the random selection of compounds is also performed.
Both the kriging and random simulations are repeated thirty times and the average number of compounds required for finding the compound with the highest melting temperature is observed.
Figure \ref{melting:kriging} shows the highest melting temperature among observed compounds during one of thirty times kriging and random trials.
As can be seen in Fig. \ref{melting:kriging}, the compound with the highest melting temperature can be found much efficiently using the kriging.
The average number of observed compounds required for finding the compounds with the highest melting temperature over thirty times trials using the kriging and random compound selections are 16.1 and 133.4, respectively, hence kriging substantially improved the efficiency of discovery.
This is a very encouraging result for application of kriging to various materials design problems.

\section{Conclusion}
In summary, we have presented applications of regression techniques to the prediction of the melting temperature of single and binary compounds. 
Prediction models are built by four kinds of regression techniques.
It is found that the SVR prediction model has the highest predictive power among the four regressions.
Also, the prediction models are much improved by considering the physical properties computed by the DFT calculation as predictor variables.
The best prediction model has been constructed by the SVR using the predictor variable set composed of elemental information and physical properties computed by the DFT calculation.
It has the CV score of 265 K and the RMS error for test data of 262 K.
In addition to the construction of prediction models, limitation of the predictive power is shown when extrapolation from training dataset is required.
We have also demonstrated simulations to find the compound with the highest melting temperature.
The simulations are based on kriging that stands on the GPR.
The average number of compounds required for finding the optimal compound over thirty-times kriging compound selection is 16.1, which are much smaller than that in random compound selections of 133.4, hence the kriging discovers the optimal compound much efficiently.
This result strongly supports that the kriging facilitates efficient discovery of optimal materials.

\begin{acknowledgments}
This study is supported by a Grant-in-Aid for Scientific Research (A) and a Grant-in-Aid for Scientific Research on Innovative Areas "Nano Informatics" (grant number 25106005) from Japan Society for the Promotion of Science (JSPS).
\end{acknowledgments}

\bibliography{melting}

\begin{thebibliography}{38}%
\makeatletter
\providecommand \@ifxundefined [1]{%
 \@ifx{#1\undefined}
}%
\providecommand \@ifnum [1]{%
 \ifnum #1\expandafter \@firstoftwo
 \else \expandafter \@secondoftwo
 \fi
}%
\providecommand \@ifx [1]{%
 \ifx #1\expandafter \@firstoftwo
 \else \expandafter \@secondoftwo
 \fi
}%
\providecommand \natexlab [1]{#1}%
\providecommand \enquote  [1]{``#1''}%
\providecommand \bibnamefont  [1]{#1}%
\providecommand \bibfnamefont [1]{#1}%
\providecommand \citenamefont [1]{#1}%
\providecommand \href@noop [0]{\@secondoftwo}%
\providecommand \href [0]{\begingroup \@sanitize@url \@href}%
\providecommand \@href[1]{\@@startlink{#1}\@@href}%
\providecommand \@@href[1]{\endgroup#1\@@endlink}%
\providecommand \@sanitize@url [0]{\catcode `\\12\catcode `\$12\catcode
  `\&12\catcode `\#12\catcode `\^12\catcode `\_12\catcode `\%12\relax}%
\providecommand \@@startlink[1]{}%
\providecommand \@@endlink[0]{}%
\providecommand \url  [0]{\begingroup\@sanitize@url \@url }%
\providecommand \@url [1]{\endgroup\@href {#1}{\urlprefix }}%
\providecommand \urlprefix  [0]{URL }%
\providecommand \Eprint [0]{\href }%
\providecommand \doibase [0]{http://dx.doi.org/}%
\providecommand \selectlanguage [0]{\@gobble}%
\providecommand \bibinfo  [0]{\@secondoftwo}%
\providecommand \bibfield  [0]{\@secondoftwo}%
\providecommand \translation [1]{[#1]}%
\providecommand \BibitemOpen [0]{}%
\providecommand \bibitemStop [0]{}%
\providecommand \bibitemNoStop [0]{.\EOS\space}%
\providecommand \EOS [0]{\spacefactor3000\relax}%
\providecommand \BibitemShut  [1]{\csname bibitem#1\endcsname}%
\let\auto@bib@innerbib\@empty
\bibitem [{\citenamefont {Curtarolo}\ \emph {et~al.}(2013)\citenamefont
  {Curtarolo}, \citenamefont {Hart}, \citenamefont {Nardelli}, \citenamefont
  {Mingo}, \citenamefont {Sanvito},\ and\ \citenamefont
  {Levy}}]{curtarolothe2013}%
  \BibitemOpen
  \bibfield  {author} {\bibinfo {author} {\bibfnamefont {S.}~\bibnamefont
  {Curtarolo}}, \bibinfo {author} {\bibfnamefont {G.~L.~W.}\ \bibnamefont
  {Hart}}, \bibinfo {author} {\bibfnamefont {M.~B.}\ \bibnamefont {Nardelli}},
  \bibinfo {author} {\bibfnamefont {N.}~\bibnamefont {Mingo}}, \bibinfo
  {author} {\bibfnamefont {S.}~\bibnamefont {Sanvito}}, \ and\ \bibinfo
  {author} {\bibfnamefont {O.}~\bibnamefont {Levy}},\ }\href {\doibase
  10.1038/nmat3568} {\bibfield  {journal} {\bibinfo  {journal} {Nature
  Materials}\ }\textbf {\bibinfo {volume} {12}},\ \bibinfo {pages} {191}
  (\bibinfo {year} {2013})}\BibitemShut {NoStop}%
\bibitem [{\citenamefont {Ceder}(2010)}]{MRS:Ceder_7945213}%
  \BibitemOpen
  \bibfield  {author} {\bibinfo {author} {\bibfnamefont {G.}~\bibnamefont
  {Ceder}},\ }\href {\doibase 10.1557/mrs2010.681} {\bibfield  {journal}
  {\bibinfo  {journal} {MRS Bulletin}\ }\textbf {\bibinfo {volume} {35}},\
  \bibinfo {pages} {693} (\bibinfo {year} {2010})}\BibitemShut {NoStop}%
\bibitem [{\citenamefont {Curtarolo}\ \emph {et~al.}(2003)\citenamefont
  {Curtarolo}, \citenamefont {Morgan}, \citenamefont {Persson}, \citenamefont
  {Rodgers},\ and\ \citenamefont {Ceder}}]{Curtarolo_PRL_2003}%
  \BibitemOpen
  \bibfield  {author} {\bibinfo {author} {\bibfnamefont {S.}~\bibnamefont
  {Curtarolo}}, \bibinfo {author} {\bibfnamefont {D.}~\bibnamefont {Morgan}},
  \bibinfo {author} {\bibfnamefont {K.}~\bibnamefont {Persson}}, \bibinfo
  {author} {\bibfnamefont {J.}~\bibnamefont {Rodgers}}, \ and\ \bibinfo
  {author} {\bibfnamefont {G.}~\bibnamefont {Ceder}},\ }\href {\doibase
  10.1103/physrevlett.91.135503} {\bibfield  {journal} {\bibinfo  {journal}
  {Phys. Rev. Lett.}\ }\textbf {\bibinfo {volume} {91}},\ \bibinfo {pages}
  {135503} (\bibinfo {year} {2003})}\BibitemShut {NoStop}%
\bibitem [{\citenamefont {Fischer}\ \emph {et~al.}(2006)\citenamefont
  {Fischer}, \citenamefont {Tibbetts}, \citenamefont {Morgan},\ and\
  \citenamefont {Ceder}}]{Fischer_NM_2006}%
  \BibitemOpen
  \bibfield  {author} {\bibinfo {author} {\bibfnamefont {C.~C.}\ \bibnamefont
  {Fischer}}, \bibinfo {author} {\bibfnamefont {K.~J.}\ \bibnamefont
  {Tibbetts}}, \bibinfo {author} {\bibfnamefont {D.}~\bibnamefont {Morgan}}, \
  and\ \bibinfo {author} {\bibfnamefont {G.}~\bibnamefont {Ceder}},\ }\href
  {\doibase 10.1038/nmat1691} {\bibfield  {journal} {\bibinfo  {journal}
  {Nature Materials}\ }\textbf {\bibinfo {volume} {5}},\ \bibinfo {pages}
  {641–646} (\bibinfo {year} {2006})}\BibitemShut {NoStop}%
\bibitem [{\citenamefont {Hautier}\ \emph {et~al.}(2012)\citenamefont
  {Hautier}, \citenamefont {Jain},\ and\ \citenamefont
  {Ong}}]{Hautier_2012:10670628}%
  \BibitemOpen
  \bibfield  {author} {\bibinfo {author} {\bibfnamefont {G.}~\bibnamefont
  {Hautier}}, \bibinfo {author} {\bibfnamefont {A.}~\bibnamefont {Jain}}, \
  and\ \bibinfo {author} {\bibfnamefont {S.~P.}\ \bibnamefont {Ong}},\ }\href
  {\doibase 10.1007/s10853-012-6424-0} {\bibfield  {journal} {\bibinfo
  {journal} {J.\ Mater.\ Sci.}\ }\textbf {\bibinfo {volume} {47}},\ \bibinfo
  {pages} {7317} (\bibinfo {year} {2012})}\BibitemShut {NoStop}%
\bibitem [{\citenamefont
  {Rajan}(2008)}]{doi:10.1146/annurev.matsci.38.060407.130217}%
  \BibitemOpen
  \bibfield  {author} {\bibinfo {author} {\bibfnamefont {K.}~\bibnamefont
  {Rajan}},\ }\href {\doibase 10.1146/annurev.matsci.38.060407.130217}
  {\bibfield  {journal} {\bibinfo  {journal} {Annu. Rev. Mater. Res.}\ }\textbf
  {\bibinfo {volume} {38}},\ \bibinfo {pages} {299} (\bibinfo {year}
  {2008})}\BibitemShut {NoStop}%
\bibitem [{\citenamefont {Yu}\ and\ \citenamefont
  {Zunger}(2012)}]{PhysRevLett.108.068701}%
  \BibitemOpen
  \bibfield  {author} {\bibinfo {author} {\bibfnamefont {L.}~\bibnamefont
  {Yu}}\ and\ \bibinfo {author} {\bibfnamefont {A.}~\bibnamefont {Zunger}},\
  }\href {\doibase 10.1103/PhysRevLett.108.068701} {\bibfield  {journal}
  {\bibinfo  {journal} {Phys. Rev. Lett.}\ }\textbf {\bibinfo {volume} {108}},\
  \bibinfo {pages} {068701} (\bibinfo {year} {2012})}\BibitemShut {NoStop}%
\bibitem [{\citenamefont {Greeley}\ \emph {et~al.}(2006)\citenamefont
  {Greeley}, \citenamefont {Jaramillo}, \citenamefont {Bonde}, \citenamefont
  {Chorkendorff},\ and\ \citenamefont {N{\o}rskov}}]{Greeley_2006}%
  \BibitemOpen
  \bibfield  {author} {\bibinfo {author} {\bibfnamefont {J.}~\bibnamefont
  {Greeley}}, \bibinfo {author} {\bibfnamefont {T.~F.}\ \bibnamefont
  {Jaramillo}}, \bibinfo {author} {\bibfnamefont {J.}~\bibnamefont {Bonde}},
  \bibinfo {author} {\bibfnamefont {I.}~\bibnamefont {Chorkendorff}}, \ and\
  \bibinfo {author} {\bibfnamefont {J.~K.}\ \bibnamefont {N{\o}rskov}},\
  }\href@noop {} {\bibfield  {journal} {\bibinfo  {journal} {Nature Materials}\
  }\textbf {\bibinfo {volume} {5}},\ \bibinfo {pages} {909} (\bibinfo {year}
  {2006})}\BibitemShut {NoStop}%
\bibitem [{\citenamefont {Castelli}\ \emph {et~al.}(2012)\citenamefont
  {Castelli}, \citenamefont {Olsen}, \citenamefont {Datta}, \citenamefont
  {Landis}, \citenamefont {Dahl}, \citenamefont {Thygesen},\ and\ \citenamefont
  {Jacobsen}}]{Castelli_C1EE02717D}%
  \BibitemOpen
  \bibfield  {author} {\bibinfo {author} {\bibfnamefont {I.~E.}\ \bibnamefont
  {Castelli}}, \bibinfo {author} {\bibfnamefont {T.}~\bibnamefont {Olsen}},
  \bibinfo {author} {\bibfnamefont {S.}~\bibnamefont {Datta}}, \bibinfo
  {author} {\bibfnamefont {D.~D.}\ \bibnamefont {Landis}}, \bibinfo {author}
  {\bibfnamefont {S.}~\bibnamefont {Dahl}}, \bibinfo {author} {\bibfnamefont
  {K.~S.}\ \bibnamefont {Thygesen}}, \ and\ \bibinfo {author} {\bibfnamefont
  {K.~W.}\ \bibnamefont {Jacobsen}},\ }\href {\doibase 10.1039/C1EE02717D}
  {\bibfield  {journal} {\bibinfo  {journal} {Energy Environ. Sci.}\ }\textbf
  {\bibinfo {volume} {5}},\ \bibinfo {pages} {5814} (\bibinfo {year}
  {2012})}\BibitemShut {NoStop}%
\bibitem [{\citenamefont {Setyawan}\ \emph {et~al.}(2011)\citenamefont
  {Setyawan}, \citenamefont {Gaume}, \citenamefont {Lam}, \citenamefont
  {Feigelson},\ and\ \citenamefont {Curtarolo}}]{Setyawan_2011}%
  \BibitemOpen
  \bibfield  {author} {\bibinfo {author} {\bibfnamefont {W.}~\bibnamefont
  {Setyawan}}, \bibinfo {author} {\bibfnamefont {R.~M.}\ \bibnamefont {Gaume}},
  \bibinfo {author} {\bibfnamefont {S.}~\bibnamefont {Lam}}, \bibinfo {author}
  {\bibfnamefont {R.~S.}\ \bibnamefont {Feigelson}}, \ and\ \bibinfo {author}
  {\bibfnamefont {S.}~\bibnamefont {Curtarolo}},\ }\href@noop {} {\bibfield
  {journal} {\bibinfo  {journal} {ACS Comb. Sci.}\ }\textbf {\bibinfo {volume}
  {13}},\ \bibinfo {pages} {382} (\bibinfo {year} {2011})}\BibitemShut
  {NoStop}%
\bibitem [{\citenamefont {Fujimura}\ \emph {et~al.}(2013)\citenamefont
  {Fujimura}, \citenamefont {Seko}, \citenamefont {Koyama}, \citenamefont
  {Kuwabara}, \citenamefont {Kishida}, \citenamefont {Shitara}, \citenamefont
  {Fisher}, \citenamefont {Moriwake},\ and\ \citenamefont
  {Tanaka}}]{Fujimura_2013_AENM:AENM201300060}%
  \BibitemOpen
  \bibfield  {author} {\bibinfo {author} {\bibfnamefont {K.}~\bibnamefont
  {Fujimura}}, \bibinfo {author} {\bibfnamefont {A.}~\bibnamefont {Seko}},
  \bibinfo {author} {\bibfnamefont {Y.}~\bibnamefont {Koyama}}, \bibinfo
  {author} {\bibfnamefont {A.}~\bibnamefont {Kuwabara}}, \bibinfo {author}
  {\bibfnamefont {I.}~\bibnamefont {Kishida}}, \bibinfo {author} {\bibfnamefont
  {K.}~\bibnamefont {Shitara}}, \bibinfo {author} {\bibfnamefont {C.~A.~J.}\
  \bibnamefont {Fisher}}, \bibinfo {author} {\bibfnamefont {H.}~\bibnamefont
  {Moriwake}}, \ and\ \bibinfo {author} {\bibfnamefont {I.}~\bibnamefont
  {Tanaka}},\ }\href {\doibase 10.1002/aenm.201300060} {\bibfield  {journal}
  {\bibinfo  {journal} {Adv. Energy Mater.}\ }\textbf {\bibinfo {volume} {3}},\
  \bibinfo {pages} {980} (\bibinfo {year} {2013})}\BibitemShut {NoStop}%
\bibitem [{\citenamefont {Lindemann}(1910)}]{Lindemann_1910}%
  \BibitemOpen
  \bibfield  {author} {\bibinfo {author} {\bibfnamefont {F.~A.}\ \bibnamefont
  {Lindemann}},\ }\href@noop {} {\bibfield  {journal} {\bibinfo  {journal}
  {Phys. Z.}\ }\textbf {\bibinfo {volume} {11}},\ \bibinfo {pages} {609}
  (\bibinfo {year} {1910})}\BibitemShut {NoStop}%
\bibitem [{\citenamefont {Guinea}\ \emph {et~al.}(1984)\citenamefont {Guinea},
  \citenamefont {Rose}, \citenamefont {Smith},\ and\ \citenamefont
  {Ferrante}}]{Guinea_1984}%
  \BibitemOpen
  \bibfield  {author} {\bibinfo {author} {\bibfnamefont {F.}~\bibnamefont
  {Guinea}}, \bibinfo {author} {\bibfnamefont {J.~H.}\ \bibnamefont {Rose}},
  \bibinfo {author} {\bibfnamefont {J.~R.}\ \bibnamefont {Smith}}, \ and\
  \bibinfo {author} {\bibfnamefont {J.}~\bibnamefont {Ferrante}},\ }\href@noop
  {} {\bibfield  {journal} {\bibinfo  {journal} {Appl. Phys. Lett.}\ }\textbf
  {\bibinfo {volume} {44}},\ \bibinfo {pages} {53} (\bibinfo {year}
  {1984})}\BibitemShut {NoStop}%
\bibitem [{\citenamefont {Van~Vechten}(1972)}]{PhysRevLett.29.769}%
  \BibitemOpen
  \bibfield  {author} {\bibinfo {author} {\bibfnamefont {J.~A.}\ \bibnamefont
  {Van~Vechten}},\ }\href {\doibase 10.1103/PhysRevLett.29.769} {\bibfield
  {journal} {\bibinfo  {journal} {Phys. Rev. Lett.}\ }\textbf {\bibinfo
  {volume} {29}},\ \bibinfo {pages} {769} (\bibinfo {year} {1972})}\BibitemShut
  {NoStop}%
\bibitem [{\citenamefont {Chelikowsky}\ and\ \citenamefont
  {Anderson}(1987)}]{Chelikowsky_1987}%
  \BibitemOpen
  \bibfield  {author} {\bibinfo {author} {\bibfnamefont {J.~R.}\ \bibnamefont
  {Chelikowsky}}\ and\ \bibinfo {author} {\bibfnamefont {K.~E.}\ \bibnamefont
  {Anderson}},\ }\href@noop {} {\bibfield  {journal} {\bibinfo  {journal} {J.
  Phys. Chem. Solids}\ }\textbf {\bibinfo {volume} {48}},\ \bibinfo {pages}
  {197} (\bibinfo {year} {1987})}\BibitemShut {NoStop}%
\bibitem [{\citenamefont {Saad}\ \emph {et~al.}(2012)\citenamefont {Saad},
  \citenamefont {Gao}, \citenamefont {Ngo}, \citenamefont {Bobbitt},
  \citenamefont {Chelikowsky},\ and\ \citenamefont
  {Andreoni}}]{PhysRevB.85.104104}%
  \BibitemOpen
  \bibfield  {author} {\bibinfo {author} {\bibfnamefont {Y.}~\bibnamefont
  {Saad}}, \bibinfo {author} {\bibfnamefont {D.}~\bibnamefont {Gao}}, \bibinfo
  {author} {\bibfnamefont {T.}~\bibnamefont {Ngo}}, \bibinfo {author}
  {\bibfnamefont {S.}~\bibnamefont {Bobbitt}}, \bibinfo {author} {\bibfnamefont
  {J.~R.}\ \bibnamefont {Chelikowsky}}, \ and\ \bibinfo {author} {\bibfnamefont
  {W.}~\bibnamefont {Andreoni}},\ }\href {\doibase 10.1103/PhysRevB.85.104104}
  {\bibfield  {journal} {\bibinfo  {journal} {Phys. Rev. B}\ }\textbf {\bibinfo
  {volume} {85}},\ \bibinfo {pages} {104104} (\bibinfo {year}
  {2012})}\BibitemShut {NoStop}%
\bibitem [{\citenamefont {Wold}\ \emph {et~al.}(2001)\citenamefont {Wold},
  \citenamefont {Sj{\"o}str{\"o}m},\ and\ \citenamefont
  {Eriksson}}]{PLS_Wold_2001}%
  \BibitemOpen
  \bibfield  {author} {\bibinfo {author} {\bibfnamefont {S.}~\bibnamefont
  {Wold}}, \bibinfo {author} {\bibfnamefont {M.}~\bibnamefont
  {Sj{\"o}str{\"o}m}}, \ and\ \bibinfo {author} {\bibfnamefont
  {L.}~\bibnamefont {Eriksson}},\ }\href@noop {} {\bibfield  {journal}
  {\bibinfo  {journal} {Chemom.\ Intell.\ Lab.\ Sys.}\ }\textbf {\bibinfo
  {volume} {58}},\ \bibinfo {pages} {109} (\bibinfo {year} {2001})}\BibitemShut
  {NoStop}%
\bibitem [{\citenamefont {Vinzi}\ and\ \citenamefont
  {G}(2013)}]{PLS_VinZi_2013}%
  \BibitemOpen
  \bibfield  {author} {\bibinfo {author} {\bibfnamefont {V.~E.}\ \bibnamefont
  {Vinzi}}\ and\ \bibinfo {author} {\bibfnamefont {R.}~\bibnamefont {G}},\
  }\href@noop {} {\bibfield  {journal} {\bibinfo  {journal} {WIREs Comput.\
  Stat.}\ }\textbf {\bibinfo {volume} {5}},\ \bibinfo {pages} {1} (\bibinfo
  {year} {2013})}\BibitemShut {NoStop}%
\bibitem [{\citenamefont {Mevik}\ and\ \citenamefont
  {Wehrens}(2007)}]{PLS_Mevik_2007}%
  \BibitemOpen
  \bibfield  {author} {\bibinfo {author} {\bibfnamefont {B.}~\bibnamefont
  {Mevik}}\ and\ \bibinfo {author} {\bibfnamefont {R.}~\bibnamefont
  {Wehrens}},\ }\href@noop {} {\bibfield  {journal} {\bibinfo  {journal} {J.\
  Stat.\ Softw.}\ }\textbf {\bibinfo {volume} {18}},\ \bibinfo {pages} {1}
  (\bibinfo {year} {2007})}\BibitemShut {NoStop}%
\bibitem [{\citenamefont {Vapnik}(1995)}]{SVR_Vapnik_1995}%
  \BibitemOpen
  \bibfield  {author} {\bibinfo {author} {\bibfnamefont {V.}~\bibnamefont
  {Vapnik}},\ }\href@noop {} {\emph {\bibinfo {title} {The Nature of
  Statistical Learning Theory}}}\ (\bibinfo  {publisher} {Springer, New York},\
  \bibinfo {year} {1995})\BibitemShut {NoStop}%
\bibitem [{\citenamefont {Vapnik}(1998)}]{SVR_Vapnik_1998}%
  \BibitemOpen
  \bibfield  {author} {\bibinfo {author} {\bibfnamefont {V.}~\bibnamefont
  {Vapnik}},\ }\href@noop {} {\emph {\bibinfo {title} {Statistical Learning
  Theory}}}\ (\bibinfo  {publisher} {Wiley, New York},\ \bibinfo {year}
  {1998})\BibitemShut {NoStop}%
\bibitem [{\citenamefont {Muller}\ \emph {et~al.}(2001)\citenamefont {Muller},
  \citenamefont {Mika}, \citenamefont {Ratsch}, \citenamefont {Tsuda},\ and\
  \citenamefont {Scholkopf}}]{Muller_SVM_914517}%
  \BibitemOpen
  \bibfield  {author} {\bibinfo {author} {\bibfnamefont {K.}~\bibnamefont
  {Muller}}, \bibinfo {author} {\bibfnamefont {S.}~\bibnamefont {Mika}},
  \bibinfo {author} {\bibfnamefont {G.}~\bibnamefont {Ratsch}}, \bibinfo
  {author} {\bibfnamefont {K.}~\bibnamefont {Tsuda}}, \ and\ \bibinfo {author}
  {\bibfnamefont {B.}~\bibnamefont {Scholkopf}},\ }\href {\doibase
  10.1109/72.914517} {\bibfield  {journal} {\bibinfo  {journal} {IEEE Trans.
  Neural Networks}\ }\textbf {\bibinfo {volume} {12}},\ \bibinfo {pages} {181}
  (\bibinfo {year} {2001})}\BibitemShut {NoStop}%
\bibitem [{\citenamefont {Bishop}(2006)}]{PRML_Bishop_2006}%
  \BibitemOpen
  \bibfield  {author} {\bibinfo {author} {\bibfnamefont {C.~M.}\ \bibnamefont
  {Bishop}},\ }\href@noop {} {\emph {\bibinfo {title} {Pattern Recognition and
  Machine Learning}}}\ (\bibinfo  {publisher} {Springer, New York},\ \bibinfo
  {year} {2006})\BibitemShut {NoStop}%
\bibitem [{\citenamefont {Chang}\ and\ \citenamefont
  {Lin}(2011)}]{LIBSVM_2011}%
  \BibitemOpen
  \bibfield  {author} {\bibinfo {author} {\bibfnamefont {C.~C.}\ \bibnamefont
  {Chang}}\ and\ \bibinfo {author} {\bibfnamefont {C.~J.}\ \bibnamefont
  {Lin}},\ }\href@noop {} {\bibfield  {journal} {\bibinfo  {journal} {ACM
  Trans. Intell. Syst. Technol.}\ }\textbf {\bibinfo {volume} {2}},\ \bibinfo
  {pages} {27} (\bibinfo {year} {2011})}\BibitemShut {NoStop}%
\bibitem [{\citenamefont {Rasmussen}\ and\ \citenamefont
  {Williams}(2006)}]{Rasmussen_2006}%
  \BibitemOpen
  \bibfield  {author} {\bibinfo {author} {\bibfnamefont {C.~E.}\ \bibnamefont
  {Rasmussen}}\ and\ \bibinfo {author} {\bibfnamefont {C.~K.~I.}\ \bibnamefont
  {Williams}},\ }\href@noop {} {\emph {\bibinfo {title} {Gaussian Processes for
  Machine Learning}}}\ (\bibinfo  {publisher} {MIT Press, Cambridge},\ \bibinfo
  {year} {2006})\BibitemShut {NoStop}%
\bibitem [{\citenamefont {Jones}(2001)}]{jones01}%
  \BibitemOpen
  \bibfield  {author} {\bibinfo {author} {\bibfnamefont {D.}~\bibnamefont
  {Jones}},\ }\href@noop {} {\bibfield  {journal} {\bibinfo  {journal} {J.
  Global Optim.}\ }\textbf {\bibinfo {volume} {21}},\ \bibinfo {pages} {345}
  (\bibinfo {year} {2001})}\BibitemShut {NoStop}%
\bibitem [{\citenamefont {Haynes}(2012)}]{CRC-Handbook-Chemistry-Physics}%
  \BibitemOpen
  \bibfield  {author} {\bibinfo {author} {\bibfnamefont {W.~M.}\ \bibnamefont
  {Haynes}},\ }\href@noop {} {\emph {\bibinfo {title} {CRC Handbook of
  Chemistry and Physics}}},\ \bibinfo {edition} {92nd}\ ed.\ (\bibinfo
  {publisher} {CRC Press},\ \bibinfo {year} {2012})\BibitemShut {NoStop}%
\bibitem [{\citenamefont {Bl{\"o}chl}(1994)}]{PAW1}%
  \BibitemOpen
  \bibfield  {author} {\bibinfo {author} {\bibfnamefont {P.~E.}\ \bibnamefont
  {Bl{\"o}chl}},\ }\href@noop {} {\bibfield  {journal} {\bibinfo  {journal}
  {Phys. Rev. B}\ }\textbf {\bibinfo {volume} {50}},\ \bibinfo {pages} {17953}
  (\bibinfo {year} {1994})}\BibitemShut {NoStop}%
\bibitem [{\citenamefont {Kresse}\ and\ \citenamefont {Joubert}(1999)}]{PAW2}%
  \BibitemOpen
  \bibfield  {author} {\bibinfo {author} {\bibfnamefont {G.}~\bibnamefont
  {Kresse}}\ and\ \bibinfo {author} {\bibfnamefont {D.}~\bibnamefont
  {Joubert}},\ }\href@noop {} {\bibfield  {journal} {\bibinfo  {journal} {Phys.
  Rev. B}\ }\textbf {\bibinfo {volume} {59}},\ \bibinfo {pages} {1758}
  (\bibinfo {year} {1999})}\BibitemShut {NoStop}%
\bibitem [{\citenamefont {Perdew}\ \emph {et~al.}(1996)\citenamefont {Perdew},
  \citenamefont {Burke},\ and\ \citenamefont {Ernzerhof}}]{GGA:PBE96}%
  \BibitemOpen
  \bibfield  {author} {\bibinfo {author} {\bibfnamefont {J.~P.}\ \bibnamefont
  {Perdew}}, \bibinfo {author} {\bibfnamefont {K.}~\bibnamefont {Burke}}, \
  and\ \bibinfo {author} {\bibfnamefont {M.}~\bibnamefont {Ernzerhof}},\
  }\href@noop {} {\bibfield  {journal} {\bibinfo  {journal} {Phys. Rev. Lett.}\
  }\textbf {\bibinfo {volume} {77}},\ \bibinfo {pages} {3865} (\bibinfo {year}
  {1996})}\BibitemShut {NoStop}%
\bibitem [{\citenamefont {Kresse}\ and\ \citenamefont {Hafner}(1993)}]{VASP1}%
  \BibitemOpen
  \bibfield  {author} {\bibinfo {author} {\bibfnamefont {G.}~\bibnamefont
  {Kresse}}\ and\ \bibinfo {author} {\bibfnamefont {J.}~\bibnamefont
  {Hafner}},\ }\href@noop {} {\bibfield  {journal} {\bibinfo  {journal} {Phys.
  Rev. B}\ }\textbf {\bibinfo {volume} {47}},\ \bibinfo {pages} {558} (\bibinfo
  {year} {1993})}\BibitemShut {NoStop}%
\bibitem [{\citenamefont {Kresse}\ and\ \citenamefont
  {Furthm{\"u}ller}(1996)}]{VASP2}%
  \BibitemOpen
  \bibfield  {author} {\bibinfo {author} {\bibfnamefont {G.}~\bibnamefont
  {Kresse}}\ and\ \bibinfo {author} {\bibfnamefont {J.}~\bibnamefont
  {Furthm{\"u}ller}},\ }\href@noop {} {\bibfield  {journal} {\bibinfo
  {journal} {Phys. Rev. B}\ }\textbf {\bibinfo {volume} {54}},\ \bibinfo
  {pages} {11169} (\bibinfo {year} {1996})}\BibitemShut {NoStop}%
\bibitem [{\citenamefont {Venables}\ and\ \citenamefont
  {Ripley}(2002)}]{Venables_MASS}%
  \BibitemOpen
  \bibfield  {author} {\bibinfo {author} {\bibfnamefont {W.~N.}\ \bibnamefont
  {Venables}}\ and\ \bibinfo {author} {\bibfnamefont {B.~D.}\ \bibnamefont
  {Ripley}},\ }\href {http://www.stats.ox.ac.uk/pub/MASS4} {\emph {\bibinfo
  {title} {Modern Applied Statistics with S}}},\ \bibinfo {edition} {4th}\ ed.\
  (\bibinfo  {publisher} {Springer},\ \bibinfo {address} {New York},\ \bibinfo
  {year} {2002})\BibitemShut {NoStop}%
\bibitem [{\citenamefont {Akaike}(1973)}]{akaike1973information}%
  \BibitemOpen
  \bibfield  {author} {\bibinfo {author} {\bibfnamefont {H.}~\bibnamefont
  {Akaike}},\ }in\ \href@noop {} {\emph {\bibinfo {booktitle} {Second
  international symposium on information theory}}}\ (\bibinfo {organization}
  {Akademinai Kiado},\ \bibinfo {year} {1973})\ pp.\ \bibinfo {pages}
  {267--281}\BibitemShut {NoStop}%
\bibitem [{\citenamefont {Meyer}\ \emph {et~al.}(2012)\citenamefont {Meyer},
  \citenamefont {Dimitriadou}, \citenamefont {Hornik}, \citenamefont
  {Weingessel},\ and\ \citenamefont {Leisch}}]{e1071}%
  \BibitemOpen
  \bibfield  {author} {\bibinfo {author} {\bibfnamefont {D.}~\bibnamefont
  {Meyer}}, \bibinfo {author} {\bibfnamefont {E.}~\bibnamefont {Dimitriadou}},
  \bibinfo {author} {\bibfnamefont {K.}~\bibnamefont {Hornik}}, \bibinfo
  {author} {\bibfnamefont {A.}~\bibnamefont {Weingessel}}, \ and\ \bibinfo
  {author} {\bibfnamefont {F.}~\bibnamefont {Leisch}},\ }\href
  {http://CRAN.R-project.org/package=e1071} {\emph {\bibinfo {title} {e1071:
  Misc Functions of the Department of Statistics (e1071), TU Wien}}} (\bibinfo
  {year} {2012}),\ \bibinfo {note} {{R} package version 1.6-1}\BibitemShut
  {NoStop}%
\bibitem [{\citenamefont {Karatzoglou}\ \emph {et~al.}(2004)\citenamefont
  {Karatzoglou}, \citenamefont {Smola}, \citenamefont {Hornik},\ and\
  \citenamefont {Zeileis}}]{kernlab}%
  \BibitemOpen
  \bibfield  {author} {\bibinfo {author} {\bibfnamefont {A.}~\bibnamefont
  {Karatzoglou}}, \bibinfo {author} {\bibfnamefont {A.}~\bibnamefont {Smola}},
  \bibinfo {author} {\bibfnamefont {K.}~\bibnamefont {Hornik}}, \ and\ \bibinfo
  {author} {\bibfnamefont {A.}~\bibnamefont {Zeileis}},\ }\href
  {http://www.jstatsoft.org/v11/i09/} {\bibfield  {journal} {\bibinfo
  {journal} {Journal of Statistical Software}\ }\textbf {\bibinfo {volume}
  {11}},\ \bibinfo {pages} {1} (\bibinfo {year} {2004})}\BibitemShut {NoStop}%
\bibitem [{\citenamefont {Friedrich}\ \emph {et~al.}(1997)\citenamefont
  {Friedrich}, \citenamefont {Berg}, \citenamefont {Broszeit},\ and\
  \citenamefont {Berger}}]{MAWE:MAWE19970280206}%
  \BibitemOpen
  \bibfield  {author} {\bibinfo {author} {\bibfnamefont {C.}~\bibnamefont
  {Friedrich}}, \bibinfo {author} {\bibfnamefont {G.}~\bibnamefont {Berg}},
  \bibinfo {author} {\bibfnamefont {E.}~\bibnamefont {Broszeit}}, \ and\
  \bibinfo {author} {\bibfnamefont {C.}~\bibnamefont {Berger}},\ }\href
  {\doibase 10.1002/mawe.19970280206} {\bibfield  {journal} {\bibinfo
  {journal} {Mater. Sci. Eng. Technol.}\ }\textbf {\bibinfo {volume} {28}},\
  \bibinfo {pages} {59} (\bibinfo {year} {1997})}\BibitemShut {NoStop}%
\bibitem [{\citenamefont {Ambacher}(1998)}]{ambacher1998growth}%
  \BibitemOpen
  \bibfield  {author} {\bibinfo {author} {\bibfnamefont {O.}~\bibnamefont
  {Ambacher}},\ }\href@noop {} {\bibfield  {journal} {\bibinfo  {journal}
  {Journal of Physics D: Applied Physics}\ }\textbf {\bibinfo {volume} {31}},\
  \bibinfo {pages} {2653} (\bibinfo {year} {1998})}\BibitemShut {NoStop}%
\end{thebibliography}%

\appendix
\section{Melting temperatures of single and binary component solids}
Table \ref{melting:data} shows the melting temperatures of single and binary component solids in the data set.

\begingroup
\squeezetable
\begin{table*}[h]
\caption{
Melting temperatures of 248 ${\rm A}_x{\rm B}_y$ binary compounds included in the data set, quoted from Ref. \onlinecite{CRC-Handbook-Chemistry-Physics}.
}
\label{melting:data}
\begin{ruledtabular}
\begin{tabular}{cc|cc|cc|cc}
Compound & Melting temp. (K) & Compound & Melting temp. (K) & Compound & Melting temp. (K) & Compound & Melting temp. (K) \\
\hline
${\rm   Al  }$         &  933.473 & ${\rm   Cs_2O   }$ &  768.15  & ${\rm   KO_2    }$ &  808.15  & ${\rm   RbO_2   }$  & 813.15  \\
${\rm   Al_2O_3    }$ &  2327.15 & ${\rm   Cs_2S   }$ &  793.15  & ${\rm   Li  }$     &  453.65  & ${\rm   S   }$      & 388.36  \\
${\rm   Al_2S_3    }$ &  1373.15 & ${\rm   CsBr    }$ &  909.15  & ${\rm   Li_2O   }$ &  1711.15 & ${\rm   S_4N_4  }$  & 451.35  \\
${\rm   Al_2Te_3   }$ &  1168.15 & ${\rm   CsCl    }$ &  919.15  & ${\rm   Li_2S   }$ &  1645.15 & ${\rm   Sb  }$      & 903.778  \\
${\rm   Al_4C_3 }$     &  2373.15 & ${\rm   CsF }$     &  976.15  & ${\rm   Li_3N   }$ &  1086.15 & ${\rm   Sb_2O_3 }$  & 928.15  \\
${\rm   AlAs    }$     &  2013.15 & ${\rm   CsH }$     &  801.15  & ${\rm   LiBr    }$ &  823.15  & ${\rm   Sb_2S_3 }$  & 823.15  \\
${\rm   AlBr_3  }$     &  370.65  & ${\rm   CsI }$     &  905.15  & ${\rm   LiCl    }$ &  883.15  & ${\rm   Sb_2Se_3}$  & 884.15  \\
${\rm   AlCl_3  }$     &  465.75  & ${\rm   CsO_2   }$ &  705.15  & ${\rm   LiF }$     &  1121.35 & ${\rm   Sb_2Te_3}$  & 893.15  \\
${\rm   AlI_3   }$     &  461.43  & ${\rm   Ga  }$     & 302.9146 & ${\rm   LiH }$     &  965.15  & ${\rm   SbBr_3  }$  & 370.15  \\
${\rm   AlN }$         &  3273.15 & ${\rm   Ga_2O_3 }$ &  2080.15 & ${\rm   LiI }$     &  742.15  & ${\rm   SbCl_3  }$  & 346.55  \\
${\rm   AlP }$         &  2823.15 & ${\rm   Ga_2S_3 }$ &  1363.15 & ${\rm   Mg  }$     &  923.15  & ${\rm   SbF_3   }$  & 560.15  \\
${\rm   AlSb    }$     &  1338.15 & ${\rm   Ga_2Se_3}$ &  1210.15 & ${\rm   Mg_2Ge  }$ &  1390.15 & ${\rm   SbI_3   }$  & 444.15  \\
${\rm   As  }$         &  1090.15 & ${\rm   GaAs    }$ &  1511.15 & ${\rm   Mg_2Si  }$ &  1375.15 & ${\rm   Se  }$      & 493.95  \\
${\rm   As_2O_3    }$ &  587.15  & ${\rm   GaBr_3  }$ &  396.15  & ${\rm   Mg_2Sn  }$ &  1044.15 & ${\rm   SeBr_4  }$  & 396.15  \\
${\rm   As_2O_5    }$ &  1003.15 & ${\rm   GaCl_2  }$ &  445.55  & ${\rm   Mg_3As_2}$ &  1473.15 & ${\rm   SeO_2   }$  & 633.15  \\
${\rm   As_2S_3    }$ &  585.15  & ${\rm   GaCl_3  }$ &  351.05  & ${\rm   Mg_3Sb_2}$ &  1518.15 & ${\rm   SeO_3   }$  & 391.15  \\
${\rm   As_2Se_3   }$ &  650.15  & ${\rm   GaI_3   }$ &  485.15  & ${\rm   MgBr_2  }$ &  984.15  & ${\rm   Si  }$      & 1687.15  \\
${\rm   As_2Te_3   }$ &  648.15  & ${\rm   GaP }$     &  1730.15 & ${\rm   MgCl_2  }$ &  987.15  & ${\rm   Si_2I_6 }$  & 523.15  \\
${\rm   As_4S_4 }$     &  580.15  & ${\rm   GaS }$     &  1238.15 & ${\rm   MgF_2   }$ &  1536.15 & ${\rm   Si_3N_4 }$  & 2173.15  \\
${\rm   AsBr_3  }$     &  304.25  & ${\rm   GaSb    }$ &  985.15  & ${\rm   MgH_2   }$ &  600.15  & ${\rm   SiC }$      & 3103.15  \\
${\rm   AsI_3   }$     &  414.15  & ${\rm   GaSe    }$ &  1233.15 & ${\rm   MgI_2   }$ &  907.15  & ${\rm   SiI_4   }$  & 393.65  \\ 
${\rm   Ba  }$         &  1000.15 & ${\rm   GaTe    }$ &  1097.15 & ${\rm   MgO }$     &  3098.15 & ${\rm   SiS_2   }$  & 1363.15  \\
${\rm   BaBr_2 }$     &  1130.15 & ${\rm   Ge  }$     &  1211.4  & ${\rm   MgS }$     &  2499.15 & ${\rm   Sn  }$      & 505.078  \\
${\rm   BaCl_2 }$     &  1234.15 & ${\rm   GeBr_2  }$ &  395.15  & ${\rm   Na  }$     &  370.944 & ${\rm   Sn_4P_3 }$  & 823.15  \\
${\rm   BaF_2  }$     &  1641.15 & ${\rm   GeBr_4  }$ &  299.25  & ${\rm   Na_2O   }$ &  1407.15 & ${\rm   SnBr_2  }$  & 505.15  \\
${\rm   BaH_2  }$     &  1473.15 & ${\rm   GeF_2   }$ &  383.15  & ${\rm   Na_2O_2 }$ &  948.15  & ${\rm   SnBr_4  }$  & 302.25  \\
${\rm   BaI_2  }$     &  984.15  & ${\rm   GeI_2   }$ &  701.15  & ${\rm   Na_2S   }$ &  1445.15 & ${\rm   SnCl_2  }$  & 520.15  \\
${\rm   BaO }$         &  2246.15 & ${\rm   GeI_4   }$ &  419.15  & ${\rm   NaBr    }$ &  1020.15 & ${\rm   SnF_2   }$  & 488.15  \\
${\rm   BaS }$         &  2500.15 & ${\rm   GeO_2   }$ &  1389.15 & ${\rm   NaCl    }$ &  1073.85 & ${\rm   SnI_4   }$  & 675.15  \\
${\rm   BaSe    }$     &  2053.15 & ${\rm   GeS }$     &  931.15  & ${\rm   NaF }$     &  1269.15 & ${\rm   SnO }$      & 1250.15  \\
${\rm   BaSi_2 }$     &  1453.15 & ${\rm   GeS_2   }$ &  1113.15 & ${\rm   NaH }$     &  911.15  & ${\rm   SnO_2   }$  & 1903.15  \\
${\rm   Be  }$         &  1560.15 & ${\rm   GeSe    }$ &  948.15  & ${\rm   NaI }$     &  934.15  & ${\rm   SnP }$      & 813.15  \\
${\rm   Be_2C  }$     &  2400.15 & ${\rm   GeTe    }$ &  997.15  & ${\rm   NaO_2   }$ &  825.15  & ${\rm   SnS }$      & 1154.15  \\
${\rm   Be_3N_2    }$ &  2473.15 & ${\rm   I_2 }$     &  386.85  & ${\rm   P   }$     &  883.15  & ${\rm   SnSe    }$  & 1134.15  \\
${\rm   BeBr_2 }$     &  781.15  & ${\rm   I_2O_4  }$ &  403.15  & ${\rm   P_2I_4  }$ &  398.65  & ${\rm   SnSe_2  }$  & 923.15  \\
${\rm   BeCl_2 }$     &  688.15  & ${\rm   IBr }$     &  313.15  & ${\rm   P_2O_5  }$ &  835.15  & ${\rm   SnTe    }$  & 1079.15  \\
${\rm   BeF_2  }$     &  825.15  & ${\rm   ICl }$     &  300.53  & ${\rm   P_2S_3  }$ &  563.15  & ${\rm   SO_3    }$  & 335.35  \\
${\rm   BeI_2  }$     &  753.15  & ${\rm   In  }$     &  429.75  & ${\rm   P_2S_5  }$ &  558.15  & ${\rm   Sr  }$      & 1050.15  \\
${\rm   BeO }$         &  2851.15 & ${\rm   In_2O_3 }$ &  2185.15 & ${\rm   P_4S_3  }$ &  446.15  & ${\rm   SrBr_2  }$  & 930.15  \\
${\rm   Bi  }$         &  544.556 & ${\rm   In_2S_3 }$ &  1323.15 & ${\rm   P_4S_7  }$ &  581.15  & ${\rm   SrCl_2  }$  & 1147.15  \\
${\rm   Bi_2O_3    }$ &  1098.15 & ${\rm   In_2Se_3}$ &  933.15  & ${\rm   Pb  }$     &  600.612 & ${\rm   SrF_2   }$  & 1750.15  \\
${\rm   Bi_2O_4    }$ &  578.15  & ${\rm   In_2Te_3}$ &  940.15  & ${\rm   Pb_3O_4 }$ &  1103.15 & ${\rm   SrH_2   }$  & 1323.15  \\
${\rm   Bi_2S_3    }$ &  1050.15 & ${\rm   InAs    }$ &  1215.15 & ${\rm   PbBr_2  }$ &  644.15  & ${\rm   SrI_2   }$  & 811.15  \\
${\rm   Bi_2Te_3   }$ &  853.15  & ${\rm   InBr_3  }$ &  693.15  & ${\rm   PbCl_2  }$ &  774.15  & ${\rm   SrO }$      & 2804.15  \\
${\rm   BiBr_3  }$     &  492.15  & ${\rm   InCl    }$ &  498.15  & ${\rm   PbF_2   }$ &  1103.15 & ${\rm   SrS }$      & 2499.15  \\
${\rm   BiCl_3  }$     &  507.15  & ${\rm   InF_3   }$ &  1445.15 & ${\rm   PbI_2   }$ &  683.15  & ${\rm   SrSe    }$  & 1873.15  \\
${\rm   BiF_3   }$     &  922.15  & ${\rm   InI }$     &  637.55  & ${\rm   PbO }$     &  1160.15 & ${\rm   SrSi_2  }$  & 1373.15  \\
${\rm   BiF_5   }$     &  424.55  & ${\rm   InI_3   }$ &  480.15  & ${\rm   PbS }$     &  1386.15 & ${\rm   Te  }$      & 722.66  \\
${\rm   BiI_3   }$     &  681.75  & ${\rm   InN }$     &  1373.15 & ${\rm   PbSe    }$ &  1351.15 & ${\rm   TeCl_4  }$  & 497.15  \\
${\rm   Ca  }$         &  1115.15 & ${\rm   InP }$     &  1335.15 & ${\rm   PbTe    }$ &  1197.15 & ${\rm   TeF_4   }$  & 402.15  \\
${\rm   Ca_3N_2    }$ &  1468.15 & ${\rm   InS }$     &  965.15  & ${\rm   PI_3    }$ &  334.35  & ${\rm   TeI_4   }$  & 553.15  \\
${\rm   CaBr_2 }$     &  1015.15 & ${\rm   InSb    }$ &  797.15  & ${\rm   Rb  }$     &  312.45  & ${\rm   TeO_2   }$  & 1006.15  \\
${\rm   CaC_2  }$     &  2573.15 & ${\rm   K   }$     &  336.65  & ${\rm   Rb_2O   }$ &  778.15  & ${\rm   TeO_3   }$  & 703.15  \\
${\rm   CaCl_2 }$     &  1048.15 & ${\rm   K_2O    }$ &  1013.15 & ${\rm   Rb_2O_2 }$ &  843.15  & ${\rm   Tl  }$      & 577.15  \\
${\rm   CaF_2  }$     &  1691.15 & ${\rm   K_2O_2  }$ &  818.15  & ${\rm   Rb_2S   }$ &  698.15  & ${\rm   Tl_2O   }$  & 852.15  \\
${\rm   CaH_2  }$     &  1273.15 & ${\rm   K_2S    }$ &  1221.15 & ${\rm   Rb_2Se  }$ &  1006.15 & ${\rm   Tl_2O_3 }$  & 1107.15  \\
${\rm   CaI_2  }$     &  1056.15 & ${\rm   K_2Se   }$ &  1073.15 & ${\rm   RbBr    }$ &  965.15  & ${\rm   Tl_2S   }$  & 730.15  \\
${\rm   CaO }$         &  2886.15 & ${\rm   KBr }$     &  1007.15 & ${\rm   RbCl    }$ &  997.15  & ${\rm   TlBr    }$  & 733.15  \\
${\rm   CaS }$         &  2797.15 & ${\rm   KCl }$     &  1044.15 & ${\rm   RbF }$     &  1068.15 & ${\rm   TlCl    }$  & 704.15  \\
${\rm   CaSi    }$     &  1597.15 & ${\rm   KF  }$     &  1131.15 & ${\rm   RbH }$     &  858.15  & ${\rm   TlF }$      & 599.15  \\
${\rm   CaSi_2 }$     &  1313.15 & ${\rm   KH  }$     &  892.15  & ${\rm   RbI }$     &  929.15  & ${\rm   TlI }$      & 714.85  \\
${\rm   Cs  }$         &  301.65  & ${\rm   KI  }$     &  954.15  & ${\rm   RbN_3   }$ &  590.15  & ${\rm   TlSe    }$  & 603.15
\end{tabular}
\end{ruledtabular}
\end{table*}
\endgroup

\end{document}